\documentclass[10pt,twocolumn,letterpaper]{article}

\usepackage{cvpr}              
\usepackage[accsupp]{axessibility}  

\newcommand{\org}{\textsuperscript{*}}

\definecolor{cvprblue}{rgb}{0.21,0.49,0.74}
\usepackage[pagebackref, breaklinks, colorlinks, allcolors=cvprblue]{hyperref}

\def\paperID{*****} 
\def\confName{CVPR}
\def\confYear{2025}

\title{NTIRE 2025 Challenge on Efficient Burst HDR and Restoration: \\
Datasets, Methods, and Results}

\author{
Sangmin Lee \hspace{-2mm}
\thanks{S. Lee, E. Park, H. Park, H. Chun, Y. Kin, A. Canelo, C. Li, C. Guo, X. Jin and R. Timofte are the NTIRE 2025 chalenge organizers, while the other authors are participants in this challenge. Each team described their own method in the report. Appendix \ref{app: affiliation} contains the authors' teams and affiliations.\\
NTIRE 2025 webpage: \href{https://cvlai.net/ntire/2025}{https://cvlai.net/ntire/2025}.\\
Code: \href{https://github.com/Eve-ctr/RawFusion}{https://github.com/Eve-ctr/RawFusion}.}
~~~~~ 
Eunpil Park\org ~~~~~
Angel Canelo \org ~~~~~
Hyunhee Park\org ~~~~~
Youngjo Kim \org ~~~~~ \\
Hyung-Ju Chun \org ~~~~~
Xin Jin \org ~~~~~
Chongyi Li \org ~~~~~
Chun-Le Guo \org ~~~~~
Radu Timofte\org ~~~~~ \\
Qi Wu ~~~~
Tianheng Qiu ~~~~
Yuchun Dong ~~~~
Shenglin Ding ~~~~
Guanghua Pan ~~~~
Weiyu Zhou ~~~~ \\
Tao Hu ~~~~
Yixu Feng ~~~~
Duwei Dai ~~~~
Yu Cao ~~~~
Peng Wu ~~~~
Wei Dong ~~~~
Yanning Zhang ~~~~ \\
Qingsen Yan ~~~~
Simon J. Larsen ~~~~
Ruixuan Jiang ~~~~
Senyan Xu ~~~~
Xingbo Wang ~~~~ 
Xin Lu~~~~ \\
Marcos V. Conde ~~~~ 
Javier Abad-Hernández ~~~~
Álvaro García-Lara ~~~~ 
Daniel Feijoo ~~~~ \\
Álvaro García ~~~~
Zeyu Xiao ~~~~
Zhuoyuan Li
}

\begin{document}
\maketitle
\begin{abstract}
    This paper reviews the NTIRE 2025 Efficient Burst HDR and Restoration Challenge, which aims to advance efficient multi-frame high dynamic range (HDR) and restoration techniques. 
The challenge is based on a novel RAW multi-frame fusion dataset, comprising nine noisy and misaligned RAW frames with various exposure levels per scene. 
Participants were tasked with developing solutions capable of effectively fusing these frames while adhering to strict efficiency constraints: fewer than 30 million model parameters and a computational budget under 4.0 trillion FLOPs. 
A total of 217 participants registered, with six teams finally submitting valid solutions. 
The top-performing approach achieved a PSNR of 43.22 dB, showcasing the potential of novel methods in this domain. 
This paper provides a comprehensive overview of the challenge, compares the proposed solutions, and serves as a valuable reference for researchers and practitioners in efficient burst HDR and restoration. 

\end{abstract}

\section{Introduction} \label{sec: intro}

Raw image fusion is a crucial task in image signal processing with widespread real-world applications. Due to missing values in the Bayer inputs and their specific form, it requires slightly different handling techniques and is considered a challenging task. Multi-frame fusion is designed to obtain high-quality images by using multiple images with consecutive time series, as is the case with RAW burst denoising \cite{godard2018deep, guo2022differentiable, mildenhall2018burst, li2022efficient, rong2020burst, xia2020basis, dudhane2022burst, dudhane2023burstormer, monod2021analysis} or RAW image super-resolution \cite{kang2024burstm, dudhane2023burstormer, wronski2019handheld, wei2023towards, deudon2020highres}. Specifically, in the last decade, there have been rapid developments in deep learning techniques for image processing to tackle increasingly difficult problems, such as 
denoising \cite{DnCNN, FFDNet, CBDNet, UPI, CycleISP, abdelhamed2020ntire, li2023ntire, lin2023improving, dudhane2022burst}, 
deblurring \cite{DeepDeblur, zhang2018dynamic, tao2018scale, MIMO-UNet, MPRNet, mao2023intriguing}, 
super-resolution (SR) \cite{bhat2021deep, EDSR, RCAN, SRMD, AdaDSR, Swinir, SRGAN}, and 
high dynamic range (HDR) reconstruction \cite{eilertsen2017hdr, liu2020single, lecouat2022high,  zou2023rawhdr, perez2021ntire, lee2018deep, chen2023learning, monod2021analysis}).

Specifically, multi-exposure images contain richer information, and various methods have been proposed to restore clean images from degraded multi-frame inputs. Over the past decade, researchers have increasingly explored multi-exposure image fusion using noisy, blurry, misaligned frames with varying exposure levels \cite{zhang2024exposure, godard2018deep, guo2022differentiable, lecouat2022high, hasinoff2016burst}. Furthermore, a previous New Trends in Image Restoration and Enhancement (NTIRE) challenge \cite{zhang2024ntire} focused on the restoration and enhancement of low-light RAW images from burst and dual-exposure inputs, leading to several novel approaches and insights \cite{xing2024high, lin2024improving, yang2024crnet, dai2024learnable, chen2024bracketing}. These advancements suggest that, despite their inherent challenges, multi-image processing methods can offer significant advantages in addressing such problems and achieving more accurate results.

Inspired by previous works, we proposed a novel challenge in NTIRE 2025 to restore an RGB image by fusing multiple degraded RAW frames. In particular, compared to previous challenges, we introduced a more difficult task by incorporating a mixed noise distribution (Gaussian and Poisson), increasing the number of input frames, and applying several challenging degradation processes. 
This multi-frame fusion task has not been covered in previous NTIRE challenges, and it is expected to significantly contribute to advancements in this field. Furthermore, another key objective of this challenge is to encourage participants to develop highly efficient network architectures suitable for on-device deployment, enabling real-world applications.  

Overall, this paper presents the NTIRE 2025 Challenge on \emph{Efficient Burst HDR and Restoration}, which aims to explore efficient multi-frame fusion solutions from a novel perspective. Participants were tasked with designing an appropriate image signal processing (ISP) module to synthesize clean, aligned HDR RGB images from nine misaligned, noisy, and degraded RAW inputs. 
In the final testing phase, six teams successfully submitted valid solutions. 
This paper provides an overview of their methods and reports their performance. 
The proposed approaches highlight potential new state-of-the-art benchmarks in efficient multi-frame HDR fusion and restoration, which are expected to not only accelerate research in this field but also drive real-world industrial applications.  


This challenge is one of the NTIRE 2025 \footnote{\url{https://www.cvlai.net/ntire/2025/}} Workshop associated challenges on: 
ambient lighting normalization~\cite{ntire2025ambient}, 
reflection removal in the wild~\cite{ntire2025reflection}, 
shadow removal~\cite{ntire2025shadow}, 
event-based image deblurring~\cite{ntire2025event}, 
image denoising~\cite{ntire2025denoising}, 
XGC quality assessment~\cite{ntire2025xgc}, 
UGC video enhancement~\cite{ntire2025ugc}, 
night photography rendering~\cite{ntire2025night}, 
image super-resolution (x4)~\cite{ntire2025srx4},
real-world face restoration~\cite{ntire2025face}, 
efficient super-resolution~\cite{ntire2025esr}, 
HR depth estimation~\cite{ntire2025hrdepth}, 
efficient burst HDR and restoration~\cite{ntire2025ebhdr}, 
cross-domain few-shot object detection~\cite{ntire2025cross}, 
short-form UGC video quality assessment and enhancement~\cite{ntire2025shortugc,ntire2025shortugc_data}, 
text to image generation model quality assessment~\cite{ntire2025text}, 
day and night raindrop removal for dual-focused images~\cite{ntire2025day}, 
video quality assessment for video conferencing~\cite{ntire2025vqe}, 
low light image enhancement~\cite{ntire2025lowlight}, 
light field super-resolution~\cite{ntire2025lightfield}, 
restore any image model (RAIM) in the wild~\cite{ntire2025raim}, 
raw restoration and super-resolution~\cite{ntire2025raw}, 
and raw reconstruction from RGB on smartphones~\cite{ntire2025rawrgb}.

\section{NTIRE 2025 Efficient Burst HDR and Restoration Challenge} \label{sec: challenge}
\subsection{Overview}
The goal of this competition is to develop an efficient algorithm for burst HDR and restoration. More precisely, the task involves generating HDR images from multiple consecutive RAW frames captured at different exposure times. This requires aligning and denoising RAW frames taken over a period of time, each with varying exposure settings (and consequently different noise levels), and then efficiently fusing them.  
This challenge simulates real-world mobile photography scenarios using devices such as smartphones or DSLR cameras. The primary objective is to inspire innovations in mobile and camera applications by encouraging the development of efficient on-device image signal processing (ISP) models. A simple baseline code and model are provided for participants as a starting kit (\href{https://github.com/Eve-ctr/RawFusion}{https://github.com/Eve-ctr/RawFusion}).  

Specifically, this competition requires solving the following three key tasks:  
\begin{enumerate}
    \item \textbf{Multi-frame RAW image fusion}: The inputs consist of consecutive multi-frame RAW Bayer images.  
    \item \textbf{HDR image synthesis}: The inputs exhibit varying brightness levels, leading to color saturation. The goal is to fuse these frames into a single High Dynamic Range (HDR) image.  
    \item \textbf{Restoration}: The inputs contain various levels of noise, and the objective is to restore a clean ground truth (GT) image.  
\end{enumerate}  

Overall, designing an efficient algorithm for both multi-frame fusion and restoration remains a challenging task. 
The ultimate goal of this competition is to develop an approach suitable for on-device deployment, enabling real-world applications on mobile devices.

\subsection{Datasets}
Instead of using existing multi-frame RAW fusion datasets such as BurstSR \cite{bhat2021deep, dudhane2023burstormer, kang2024burstm, luo2022bsrt}, SyntheticBurst \cite{kang2024burstm, wei2023towards, bhat2023self, dudhane2022burst}, LLFF-N \cite{pearl2022nan, tanay2023efficient}, or other datasets from previous NTIRE challenges \cite{bhat2021ntire, bhat2022ntire, zhang2024ntire}, we introduce a novel synthetic RAW HDR fusion dataset in this competition to encourage innovative approaches.  

Specifically, the training dataset is generated using a virtual imaging pipeline. For each scene, nine clean RAW frames with various exposure levels are captured, and the corresponding ground-truth (GT) HDR RGB image is synthesized from these frames. The provided RAW frames have three different exposure levels (low, medium, and high), resulting in three distinct noise levels. To simulate real-world burst RAW acquisition, the dataset incorporates various degradations, including noise, rotation, translation, and motion blur.  

\begin{figure}[ht]
    \centering
    \includegraphics[width=0.98\linewidth]{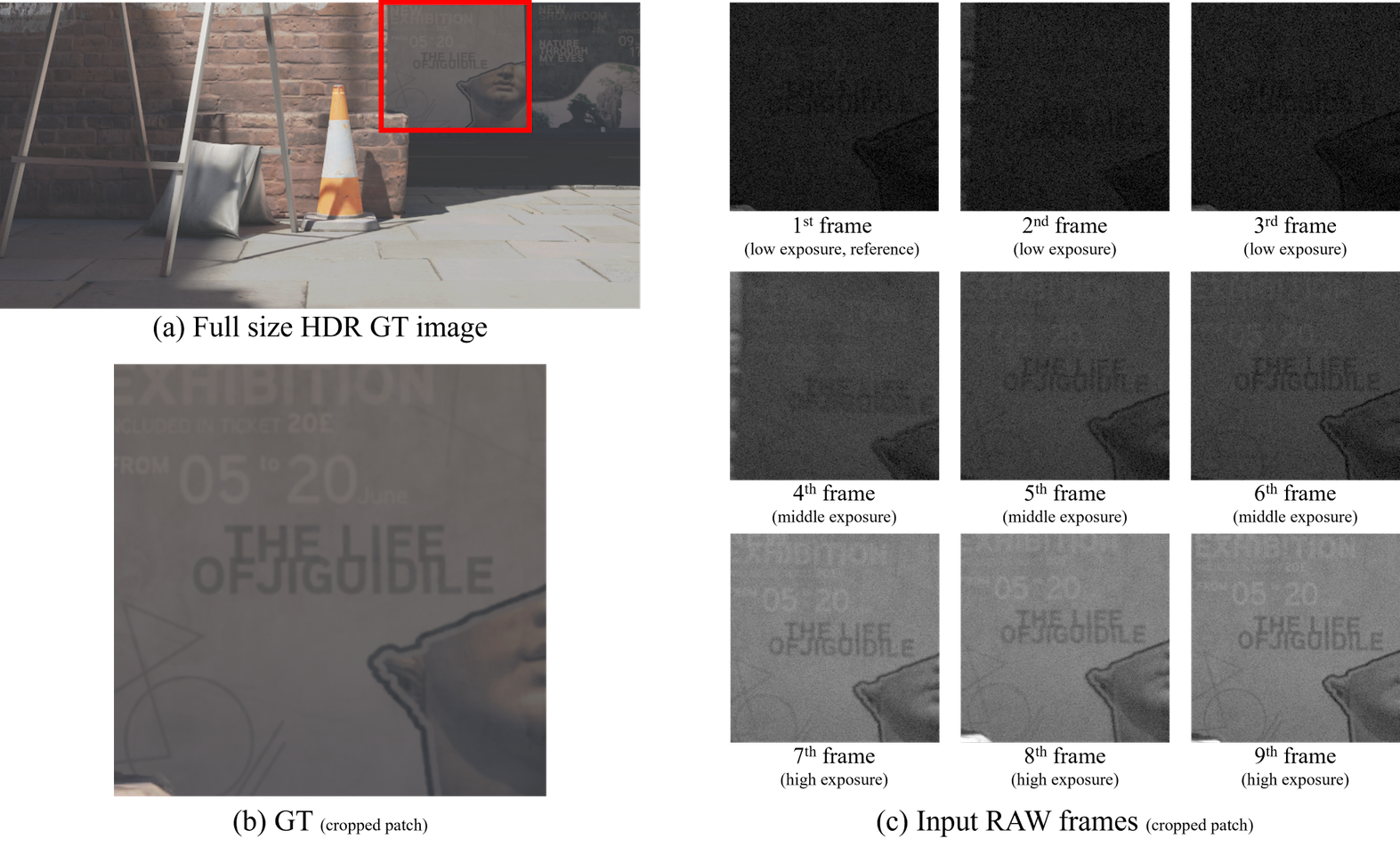}
    \caption{
    Visualization of the dataset proposed in this competition.  
    (a) Example of an HDR RGB GT image.  
    (b) Cropped patch of (a).  
    (c) Nine input RAW frames corresponding to (b), captured at three different exposure levels. Participants are required to fuse these noisy, misaligned and degraded frames to reconstruct the ground-truth RGB image in (b).  \vspace{-5mm}
    }
    \label{fig: DB}
\end{figure}

The first frame in each sequence serves as the \textit{reference frame}, which is perfectly aligned with the ground-truth image. Figure \ref{fig: DB} illustrates an example composition of a scene in the dataset. The core objective of this dataset is to encourage participants to explore \textit{how to effectively utilize the information from multiple input frames while considering the unique characteristics of each frame}.  

The training, validation, and test datasets are constructed using the same methodology, with all images having a resolution of $768(H) \times 1536(W)$, as depicted in Figure \ref{fig: DB}(a). The training set consists of 300 scenes, each containing nine RAW input frames and one ground-truth RGB image. The validation and test sets each contain 20 scenes, with the GT images in these sets hidden from participants.  

\subsection{Challenge Rules}
\subsubsection{Metric}
The evaluation is based on the comparison between the generated images and the corresponding ground-truth images. Following previous NTIRE challenges \cite{bhat2022ntire, zhang2023ntire, li2023ntire, wang2023ntire, wang-2023ntire, zhang2024ntire, chen2024ntire, wang2024ntire}, we use the average Peak Signal-to-Noise Ratio (PSNR) \cite{huynh2008scope} across the full test set to determine the final ranking. Although the dataset is provided in 16-bit \texttt{.tif} format, PSNR is measured after conversion to 8-bit format.  
As an additional reference, we also provide the Structural Similarity Index Measure (SSIM) \cite{wang2004image}, computed in grayscale, to participants.

\subsubsection{Restrictions on Model Parameters and FLOPs} \label{sec: limitation}
Since this competition emphasizes the development of efficient on-device models, there are strict constraints on computational cost and memory. Specifically, submitted models must not exceed $30M$ parameters and $4T$ FLoating point OPerations (FLOPs) when generating a full-sized output RGB image of size $1 \times 3 \times 768 \times 1536$. The FLOP count is measured using the \texttt{FlopCountAnalysis} function from the \texttt{fvcore} module, as adopted in prior works \cite{ren2024ninth, zamfir2024see}.  

\subsubsection{Others}
Participants are allowed to employ any training techniques as long as they adhere to the constraints outlined in Section \ref{sec: limitation}.  
For instance, they may utilize external public datasets for pre-training, incorporate pre-trained networks, or leverage generative models/transformers, etc.  
Furthermore, there are no further constraints on memory usage or inference time.  
Lastly, the use of all input frames is not mandatory. To optimize computational efficiency and reduce FLOPs, participants may selectively use a subset of frames before applying fusion techniques.

\subsection{Challenge Phases}
\textit{(1) Training and validation phase}:  
Participants are provided with 300 training scene pairs and 20 validation inputs from our constructed dataset.  
The training phase lasts for three weeks, during which participants can submit their validation results to receive PSNR feedback.  
The GT images for the validation set remain hidden from participants. Instead, they can submit their enhanced results to the evaluation server, which computes PSNR and SSIM scores for the generated images and provides immediate feedback. The resulting PSNR values are displayed on the leaderboard, visible for all participants.

\noindent\textit{(2) Testing phase}:  
The test phase lasts for one week, during which participants can submit their final results.  
They are given access to the testset consisting of 20 scenes, with GT images remaining undisclosed. Similar to the validation phase, PSNR scores for test results are displayed on the leaderboard.
Participants are required to submit their enhanced outputs to the Codalab evaluation server, along with an email to the organizers containing their code and a factsheet. The organizers verify and execute the provided code to verify the final results.

\section{Challenge Results} \label{sec: results}

\begin{table}
  \centering
  \resizebox{\columnwidth}{!}{
  \begin{tabular}{cccccc}
    \toprule
    Team name & PSNR$\uparrow$ & SSIM$\uparrow$ & Params (M) & FLOPs (T) & Time (s) \\
    \midrule
    ImvisionAI & \textbf{43.22} & \textbf{0.992} & 29.051 & 3.965 & 7.12 \\
    DeepTrans & 42.75 & 0.991 & 12.839 & 3.864 & 0.24 \\
    SimonLarsen & 41.38 & 0.989 & 18.651 & 1.685 & 0.74 \\
    E\_Group & 40.64 & 0.988 & 8.386 & 0.288 & 0.62 \\
    CidautAI & 37.51 & 0.979 & 3.346 & 0.279 & 0.064 \\
    X-L & 33.85 & 0.934 & 1.73 & 0.346 & 0.088 \\
    \bottomrule
  \end{tabular}
  }
  \caption{The final results of participants. 
  Team ImvisionAI achieved the highest PSNR value of 43.22 dB using maximal resources close to the competition restriction.
  The inference time represented in the rightmost column is measured by a single NVIDA H100 GPU.
  }
  \label{tab: results}
\end{table}

\begin{figure*}[ht]
    \centering
    \includegraphics[width=0.98\linewidth]{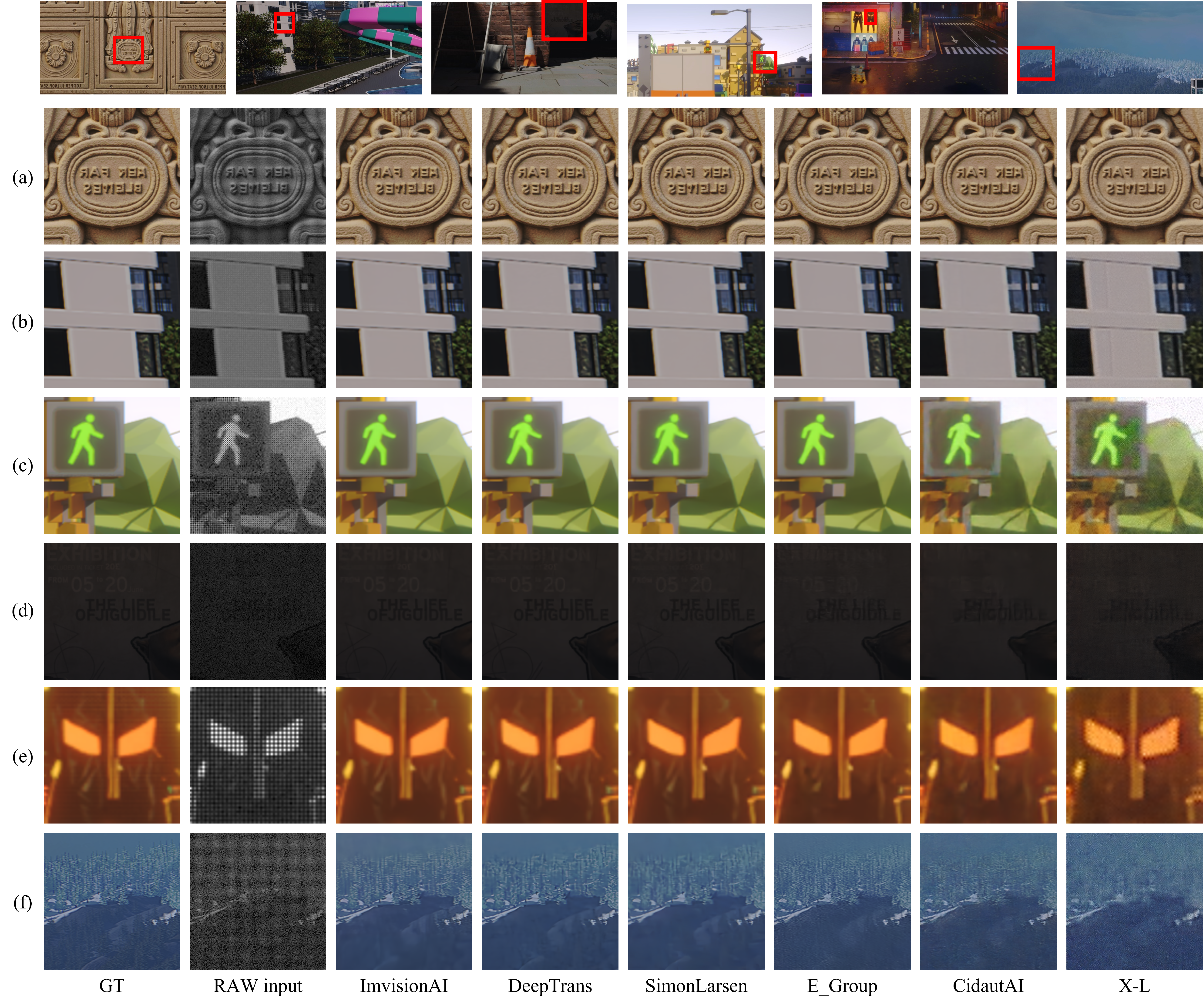}
    \caption{
    The visualization of the participants' final results, along with the corresponding input reference frames and GT images.
    The first row displays the full input scenes, while the subsequent rows show cropped patches from the images. 
    Each column represents the GT image, the input reference frame, and the results produced by the participants' models. 
    Higher-performing models generate outputs that closely restore the GT, whereas the last two rows (e, f) illustrate cases where complete restoration was unsuccessful.
    }
    \label{fig: results}
\end{figure*}

The competition had a total of 217 participants, with six teams ultimately submitting their final results. Their quantitative performance is summarized in Table \ref{tab: results}, while Figure \ref{fig: results} visualizes qualitative results for several scenes from the test set.

Team ImvisionAI won the competition, achieving the highest PSNR value of 43.22 dB on the test set, outperforming the second-place team by 0.47 dB. Their model was designed to fully utilize the competition’s constraints ($30$M model parameters and $4.0$T FLOPs, as described in Section \ref{sec: limitation}) to maximize performance. Additionally, their model also achieved the highest SSIM score of 0.992, slightly surpassing the second-place result.

\begin{figure*}[ht!] 
    \centering \includegraphics[width=0.9\textwidth]{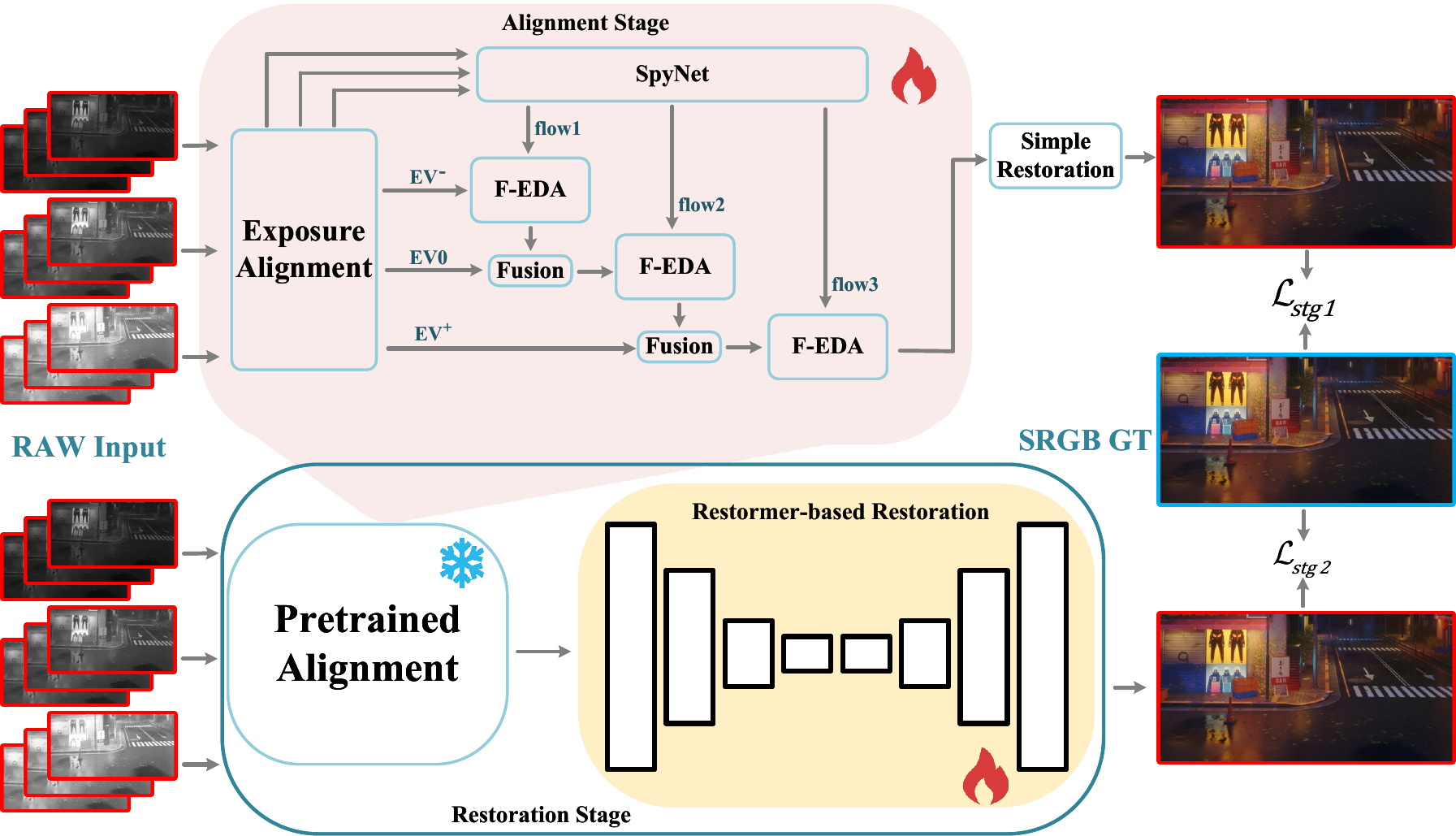} 
    \caption{
    The model proposed by the team ImvisionAI, named recursive multi-exposure alignment with spatiotemporal decoupling. 
    They designed two stages of training the models, with each stage focusing on learning multi-frame alignment and image restoration, respectively.
    After the alignment module has been trained, it is frozen during the training of the restoration model in the second stage. 
    } 
    \label{fig:ARCH} 
\end{figure*}

From an efficiency perspective, inference time for each model was measured under the same conditions using a single NVIDIA H100 GPU. The time required to restore a full-size ($768\times1536$) image was averaged and is reported in Table \ref{tab: results}. Team ImvisionAI’s winning model had the longest inference time at approximately 7 seconds, whereas team CidautAI achieved the fastest processing time at 64 ms per image. Given the trade-off between speed and performance, selecting an appropriate model depends on the specific application requirements.

Notably, an alternative approach to efficiency evaluation involves using a PSNR/time metric, as seen in other challenges \cite{conde2023efficient, ignatov2022learned}. If the competition had adopted this metric, the rankings and model designs might have differed. For instance, Team ImvisionAI’s slow inference speed is not due to GPU computation but rather an architectural bottleneck, which could be further optimized. It is important to note that this competition did not impose running time or memory usage constraints beyond those defined in Section \ref{sec: limitation}; instead, the primary focus was on achieving the best performance within the given resource limits.

To compare the qualitative performance of the participants' models, we selected six representative test scenes that highlight key characteristics of their results. These selected scenes are displayed in the first row of Figure \ref{fig: results}, with cropped patches zoomed in for detailed inspection.

In Figure \ref{fig: results}(a), all participants successfully restored textures in the bright regions of the scene. However, artifacts start to appear in the synthesized images, such as the vertical line in the rightmost column of Figure \ref{fig: results}(b) or the noisy outputs in the right two columns of (c). This issue becomes even more evident in the restoration of letters in dark areas, as shown in (d), where only the top three teams produced clearly recognizable results.

Figures \ref{fig: results}(e) and (f) illustrate two of the most challenging cases in the test set. In (e), faint horizontal lines in the GT image were not fully restored, while in (f), the details of trees in dark areas were not completely reconstructed. 
More specifically, the bright forest regions in (f) were slightly better restored by team ImvisionAI compared to team DeepTrans, whereas both struggled to recover details in the dark regions. 
Restoring dark areas is particularly difficult because the original signal is only partially available from overexposed frames. Considering the extreme noise levels in the input frames, as displayed in the second column of the figure, these cases exemplify the inherent challenges of this competition.

Finally, we analyze the results in relation to model complexity. Notably, the second-place team DeepTrans, achieved visually comparable results while using less than half the number of parameters compared to team ImvisionAI. Although some differences in fine details were observed, their approach demonstrates an effective balance between performance and efficiency, making it a strong candidate for real-world applications where time and memory constraints are critical, such as on-device environment. 

\section{Challenge Methods} \label{sec: methods}

\subsection{ImvisionAI} \label{sec: imvisionai}
\textbf{General method description.} 
The team proposed an recursive multi-exposure alignment with spatiotemporal decoupling
for efficient burst hdr and restoration named RASD~\cite{qiu2025recursive}. Burst HDR and Restoration from RAW to RGB, aiming for an end-to-end solution, is a challenging task due to two key contradictions: (1) the information differences caused by brightness, noise, saturation, and dead-black regions across multi-exposure data domains, and (2) the disparity between alignment and restoration in the temporal and spatial domains. Neglecting these factors would lead to inefficient joint optimization.

\begin{figure*}[ht!]
    \centering
    \includegraphics[width=\textwidth]{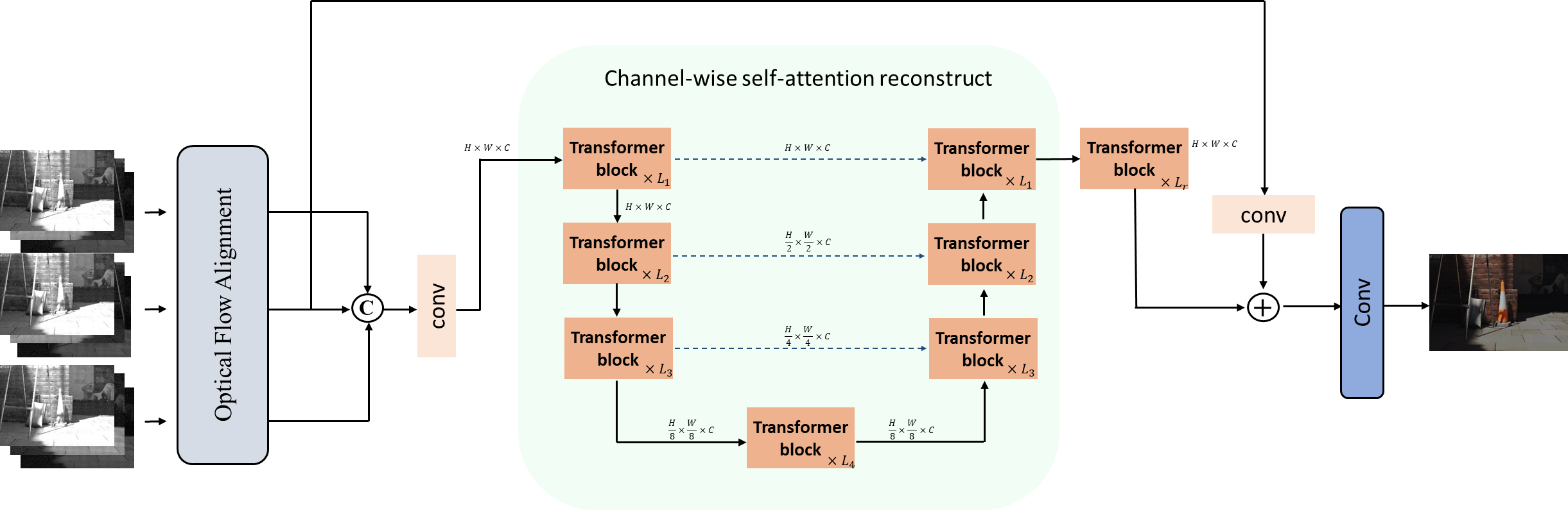}
    \caption{The proposed model by DeepTrans, named flow-guided deformable alignment with channel-wise self-attention reconstruction.}
    \label{fig:DeepTrans}
\end{figure*}

To address these challenges, the team propose a framework that explicitly considers these contradictions at both the alignment and restoration stages. Specifically, as illustrated in Fig.~\ref{fig:ARCH}, for alignment, the team leverage statistically derived luminance priors for truncation-free exposure alignment toward the longest exposure to prevent alignment errors caused by too many saturated pixels and introduce a recursive \emph{Flow-based Enhanced Deformable Alignment} (F-EDA) module to achieve multi-exposure spatial alignment. Specifically, since spynet~\cite{ranjan2017optical} is trained in the nonlinear RGB domain, the team use a learnable feature mapping to transform the input into a domain with a smaller gap, reducing flow errors and enabling more accurate optical flow estimation by spynet~\cite{ranjan2017optical}. These ensure more stable and accurate alignment across different exposure levels. Additionally, to balance the trade-off between alignment and restoration, the team adopt a two-stage training strategy that decouples spatiotemporal learning, in the align stage (stage 1), the team use a few Restormer~\cite{zamir2022restormer} blocks to maximize alignment. In the restoration stage (stage 2), the team use the full Restormer architecture to maximize the overall restoration, addressing blurring, noise, and providing a RAW to sRGB mapping. Additionally, they found that decomposing the Bayer pattern channels during the restoration phase significantly reduced accuracy. Therefore, they kept the original RAW pattern format. They analyzed that the decomposition operation lost the spatial relationships between different color channels, which lowered the restoration accuracy.

\textbf{Implementation details.} The team used AdamW optimizer ($\alpha=0.9$ and $\beta=0.9$) with 1000 training epochs each for stage 1 and
stage 2. The initial value of the learning rate was $8 \times 10^{-4}$ 
and was updated with cosine annealing schedule. The patch size was set to $128 \times 128$ pixels. For data augmentation, they only used bayeraug~\cite{liu2019learning} for random flips and transposition. In the testing phase, the team used TLC~\cite{chu2022improving} for full image inference, and note that the addition of TLC brings no complexity gain.

\subsection{DeepTrans}
\textbf{General method description.} 
The overall architecture of their model is illustrated in Figure \ref{fig:DeepTrans}. They first adopt an Optical Flow Alignment Block (OFAB) to align the nine input RAW frames. This module comprises convolutional layers for shallow feature extraction and integrates a pre-trained SpyNet \cite{ranjan2017optical} for the optical flow estimation. As the original model is designed for four-channel RAW inputs, and the input data consists of single-channel RAW images, they convert them into a four-channel format according to the RGGB Bayer pattern. This conversion reduces the spatial resolution of each frame by half.

After alignment, the nine converted frames are concatenated and passed through a transposed convolution layer to restore the original spatial resolution. This operation enables effective aggregation and utilization of complementary information from the multiple input frames.

The fused features are then fed into a four-level symmetric encoder-decoder architecture to extract deep representations. Each level of the encoder-decoder consists of multiple transformer blocks. The encoder progressively downsamples the spatial dimensions while increasing the feature dimensionality, and the decoder correspondingly upsamples the low-resolution latent features to reconstruct high-resolution representations. To facilitate the reconstruction process, skip connections are introduced to concatenate encoder features with corresponding decoder features at each level.

A convolutional layer is subsequently applied to the refined features to predict the residual image. Finally, a transposed convolution layer \cite{zeiler2011adaptive} is employed to generate an aligned reference frame with the same number of channels as the residual. The restored image is obtained by adding the residual to this reference frame.

\begin{figure*}[!ht]
    \centering
    \includegraphics[width=\textwidth]{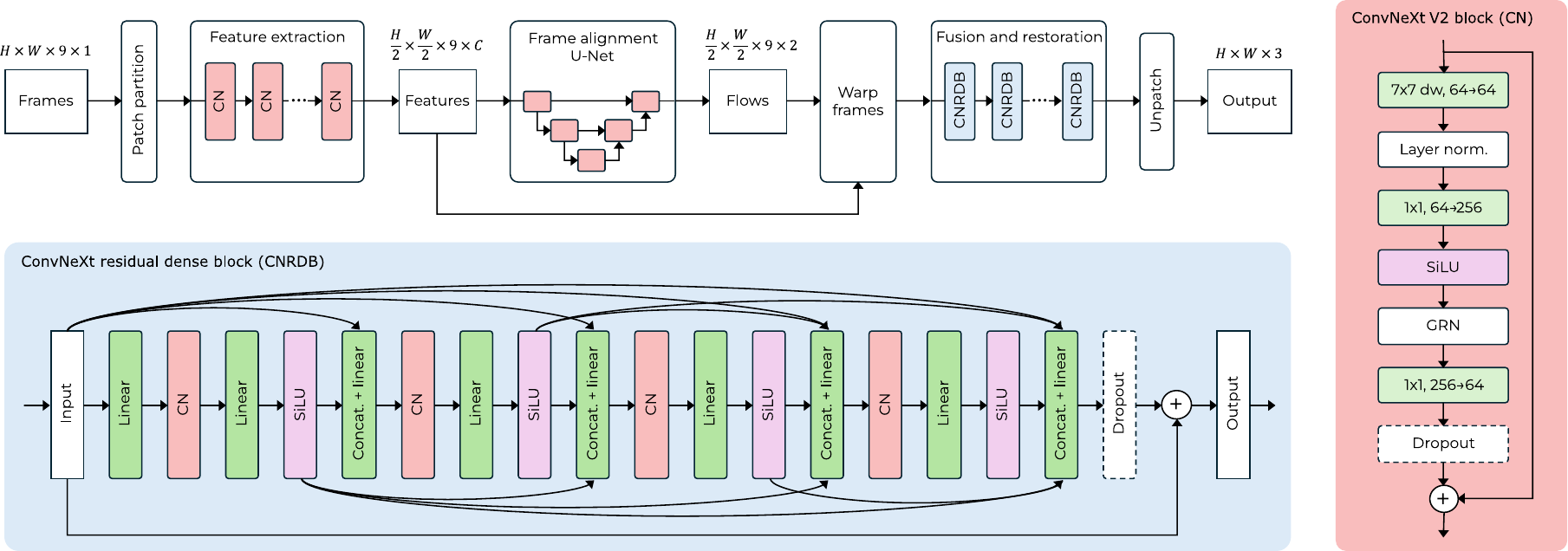}
    \caption{The model proposed by Team SimonLarsen. The full three-stage model architecture is shown in the top, the basic CN block on the right and the CNRDB block in the bottom. \emph{dw} in the CN block denotes depth-wise convolution.\vspace{-2mm}
    }
    \label{fig:SimonLarsen_architecture}
\end{figure*}

\textbf{Implementation details.} During the training process, images are randomly cropped to patches with a size of \(128 \times 128\), and are subject to random flipping and rotation. They used the PyTorch framework and adopted the Adam optimizer with $\beta_1$ = 0.9 and $\beta_2$ = 0.999. The initial learning rate was set to $2\times10^{-4}$, using CosineAnnealingLR \cite{loshchilov2022sgdr} that gradually reduces to $2\times10^{-6}$ over 80 epochs. The model was trained on 8 RTX 4090 GPUs for a total of 80 epochs, amounting to 1 day of training. 

\subsection{SimonLarsen}
\paragraph{General method description.}
Team SimonLarsen proposed the FlowCNRDB model shown in Figure~\ref{fig:SimonLarsen_architecture}.
They aimed to build a modern fusion and restoration model without self-attention by instead applying the design principles outlined in ConvNeXt \cite{liu2022convnet} and ConvNeXt V2 \cite{woo2023convnext} to produce an efficient and fully-convolutional model.
The overall network architecture consists of three stages: feature extraction, frame alignment, and finally a joint fusion and restoration model. In the first and second stages, each frame is processed in parallel, while the last stage processes all frames together to produce the final output.

All stages of the model utilized the basic block design of ConvNeXt V2 employing large kernel depth-wise convolutions followed by a point-wise MLP and Global Response Normalization (GRN). To avoid overfitting, dropout is used throughout the model.
The global feature aggregation in GRN will lead to train-test time inconsistency when trained on patched inputs. To mitigate this issue, a local aggregation method similar to test-time local conversion \cite{chu2022improving} was used during inference.

The \textbf{feature extraction stage} encodes each frame $\mathbf{I}_i \in \mathbb{R}^{H \times W \times 1}$ into a latent representation $\mathbf{Z}_i \in \mathbb{R}^{\frac{H}{2} \times \frac{W}{2} \times C}$. The input frames are first patched using a $2 \times 2$ convolution with stride 2. The patch size was chosen to match the GRBG filter size of the raw Bayer inputs. Deep features are then extracted through a series of ConvNeXt (CN) blocks.

The \textbf{frame alignment} model is a U-Net architecture \cite{ronneberger2015unet} consisting of CN blocks. Given an encoded reference and target frame $(\mathbf{Z}_\text{ref}, \mathbf{Z}_i) \in \mathbb{R}^{\frac{H}{2} \times \frac{W}{2} \times 2C}$ (concatenated on the channel axis), the model outputs a dense flow field $\mathbf{F}_i \in \mathbb{R}^{\frac{H}{2} \times \frac{W}{2} \times 2}$ estimating the displacement between the reference and the target.

The predicted flow field is used to warp the target frame to match the reference frame using bilinear grid sampling. Tokens outside the target frame are padded with zeros, and a padding mask is computed to be used in the fusion stage.

The \textbf{fusion and restoration} stage takes in the full set of aligned frames and padding masks, concatenated on the channel axis, and predicts the final RGB output image $\hat{\textbf{Y}} \in \mathbb{R}^{H \times W \times 3}$.
During training, frames are randomly shuffled within the three exposure time groups to effectively prevent the model from associating specific channels with the reference frame. This step is necessary to force the model to utilize all frames and, as a result, learn how to align the frames in the second stage.

\begin{figure*}[!ht]
    \centering
    \includegraphics[width=\textwidth]{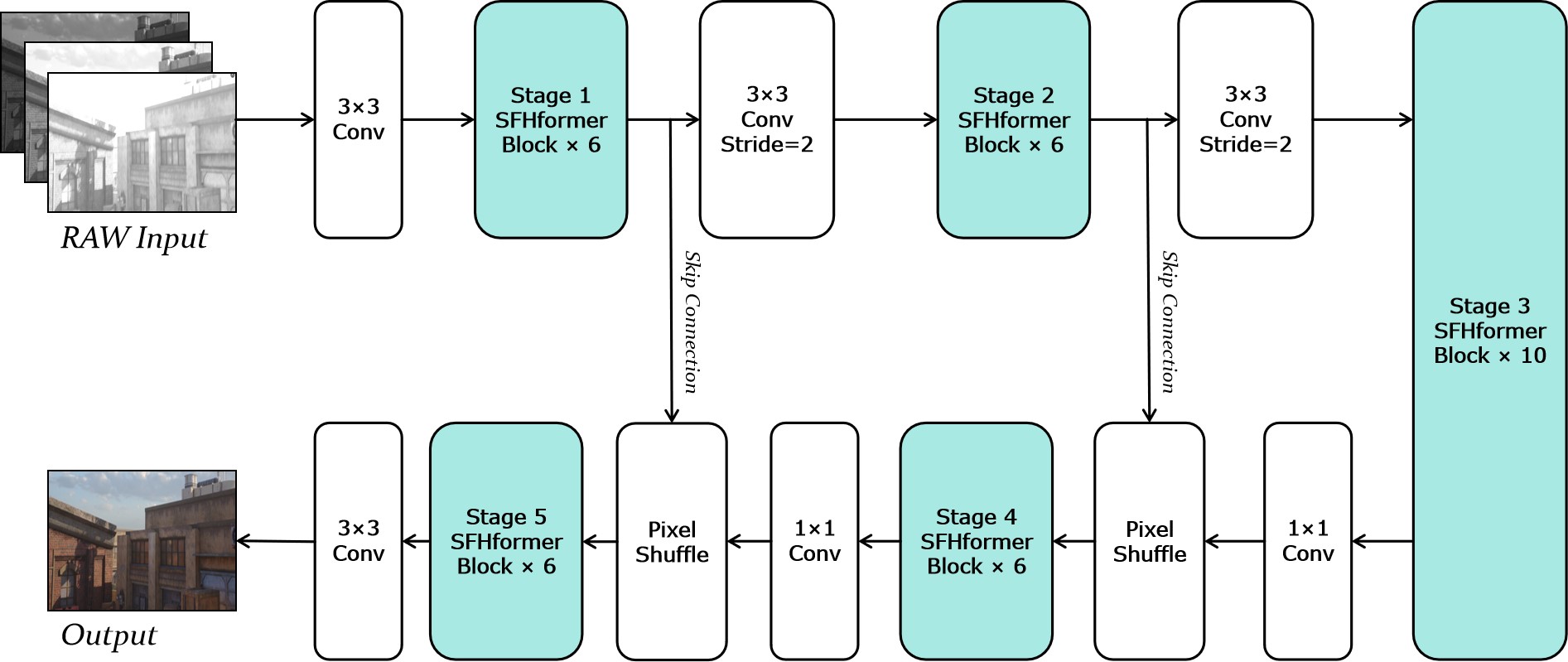}
    \caption{The proposed model by team E\_Group.}
    \label{fig:E_Group}
\end{figure*}

The fusion stage consists of a series of ConvNeXt Residual Dense Blocks (CNRDB) using a macro design similar to DRCT \cite{hsu2024drct}, but replacing Swin transformer layers with CN blocks. In addition, a linear layer was used to project the number of channels down to a common size before each CN block to constrain the number of parameters. The output from the fusion stage is unpatched with a $2 \times 2$ pixel shuffle operation followed by a $3 \times 3$ convolution down to 3 channels to produce the final output image.
\vspace{-5mm}
\paragraph{Training details.}
The model was trained solely on the provided training dataset with the last 20 scenes reserved for validation. For data augmentation, Bayer-preserving \cite{liu2019learning} horizontal/vertical flipping, rotation (in 90 degree increments) and random cropping was used. The model was trained solely with MSE loss. No additional losses were used to supervise the frame alignment.

The model was trained at $256 \times 256$ resolution with batch size 10 for 224k iterations on a single NVIDIA RTX A6000 GPU using automatic mixed precision.
The initial learning rate was set at $3 \times 10^{-4}$ and gradually reduced to $10^{-7}$ using a cosine annealing schedule. Throughout training, the AdamW optimizer was used with $\text{weight decay} = 0.01$, $\beta_1 = 0.9$, and $\beta_2 = 0.999$.

\subsection{E\_Group}

\textbf{General method description.} Team E\_Group proposed a model that utilizes the SFHFormer \cite{jiang2024} block, which consists of a hierarchical encoder-decoder structure composed of five stages. The structure includes a two-scale encoder (stage-1 and stage-2), a bottleneck (stage-3), and a two-scale decoder (stage-4 and stage-5). The core components of the SFHFormer block are: (a) Local Global Perception Mixer (LGPM) and (b) Multi-kernel ConvFFN (MCFN). The LGPM is designed to capture both local and global feature representations, while the MCFN enhances the feature transformation capabilities through multiple kernel convolutions.

\textbf{Training strategy.} Team E\_Group employs the AdamW optimizer with $\beta_1=0.9$ and $\beta_2=0.999$ to train the SFHFormer model. The initial learning rate was set to $7.5 \times 10^{-4}$, and a cosine annealing strategy \cite{loshchilov2022sgdr} was adopted to gradually decrease the learning rate from the initial value to $5 \times 10^{-6}$. This learning rate scheduler is crucial for fine-tuning, as it enables smooth convergence and helps the model escape local minima by periodically restarting the annealing process.

In addition to the optimizer settings, Team E\_Group's training process integrated mixup data augmentation with a mixup parameter $\beta=1.2$ and identity mapping enabled. This augmentation strategy enhances data diversity, leading to improved model robustness by interpolating between training samples.

Furthermore, Team E\_Group's framework utilized the FFT loss (fftLoss) for pixel-level optimization. The FFT loss, with a weight of 1.0 and mean reduction, operates in the frequency domain to capture high-frequency details, thereby improving the recovery of fine textures and edges in the restored images. This approach mitigates the blurring issues commonly associated with direct pixel-wise losses such as L1 or L2.

Team E\_Group also incorporated pre-trained weights into the training pipeline. By initializing the model with weights obtained from previous training stages, the fine-tuning process was accelerated and the model benefits from the learned feature representations, ensuring better performance in the restoration task.

All experiments were conducted using PyTorch 1.11.0 on a system equipped with two NVIDIA 4090 GPUs. The batch size was set to 2 and a crop size of 256 is used, which were optimized for both memory efficiency and effective training on high-resolution image data. This comprehensive training strategy, combining advanced optimizer settings, dynamic learning rate scheduling, effective data augmentation, and frequency-domain loss computation, ensures that the SFHFormer model achieves a balanced performance in both global structure recovery and local detail enhancement.

\subsection{CidautAI}
\begin{figure*}[!ht]
  \centering
  \includegraphics[width=\linewidth]{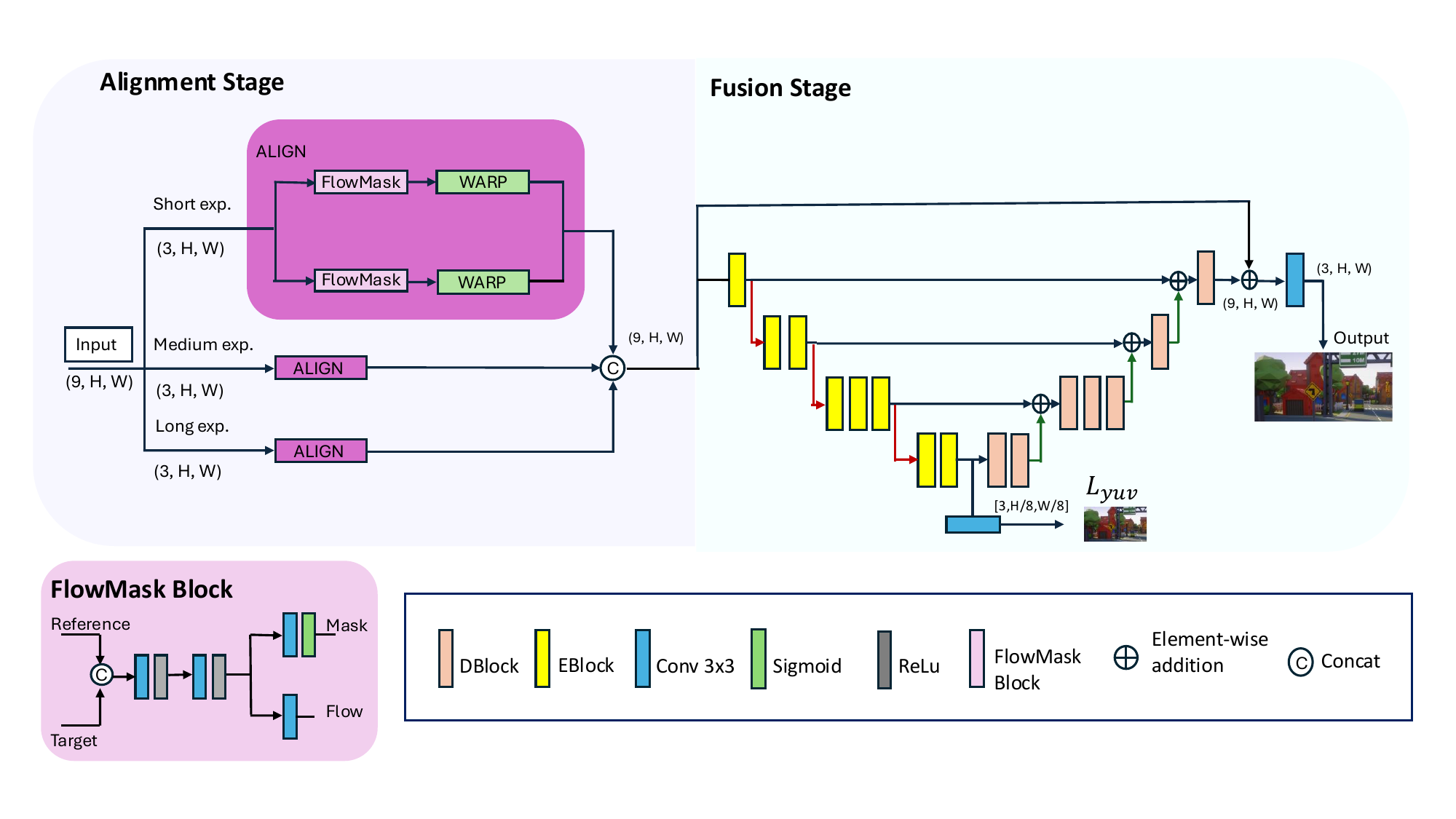}
  \caption{General diagram of \textbf{DarkIR-Fusion} proposed by team CidautAI.}
  \label{fig:darkir-fusion}
\end{figure*}

\textbf{General method description.} 
The challenge involves handling sequences of raw images with varying exposures, aligning them, and then fusing them into a single HDR image, all while minimizing computational resources. {Team CidautAI proposed the most efficient solution in terms of FLOPs and runtime.} Their solution, named DarkIR-Fusion, addresses these constraints through a highly optimized method. The solution is based in the good results of DarkIR~\cite{feijoo2024darkir} in Low-Light Image Enhancement (LLIE).

The architecture of their proposed model, \textbf{DarkIR-Fusion}, is presented in Figure \ref{fig:darkir-fusion}. During the first stage, the set of nine input frames of different exposure times are passed through an Alignment stage. Then, in the second stage, the aligned images are introduced into the encoder-decoder DarkIR architecture, which finally returns a fused HDR image of the input set. The method is $\mathbf{10\times}$ smaller than the requested size (30M), requires $\mathbf{14\times}$ fewer operations than the hardware requirements, and runs in real time ($\approx$ \textbf{15} FPS ) on an NVIDIA H100 GPU --- achieving almost 8 FPS on an RTX 4090.
\vspace{-4mm}
\paragraph{Alignment stage.} The nine input images are introduced and are grouped into the 3 trios of same exposition given in the dataset (see Figure \ref{fig:darkir-fusion}). Each trio is introduced into the Alignment Module, and for each of the images the optical flow (flow) and mask are estimated using the FlowMask Block. This block applies cascaded convolutions to the concatenation of the reference and target images, returning the Mask and the Flow (see Figure~\ref{fig:darkir-fusion}).
\vspace{-4mm}
\paragraph{DarkIR.} The neural network follows an encoder-decoder architecture based on the Metaformer structure \cite{yu2022metaformer}. The encoder focuses on the HDR illumination and color issues using Fourier information. Thus, the encoder produces a low-resolution fused image $\hat{x}_{\downarrow8}$ with corrected illumination and color. The team ensures that the encoder is working in these issues by the introduction of a guiding loss. The decoder focuses on upscaling and reducing the misalignment blur and noise artifacts using the prior illumination-enhanced encoded features. The output of this DarkIR implementation is built upon a residual connection with the output of the Alignment Stage. This [9,H,W] tensor is finally transformed into the output image using a convolutional 3x3 layer --- which can be understood as a fusion-demosaicing layer.

\begin{figure*}[!ht]
    \centering
    \includegraphics[width=\textwidth]{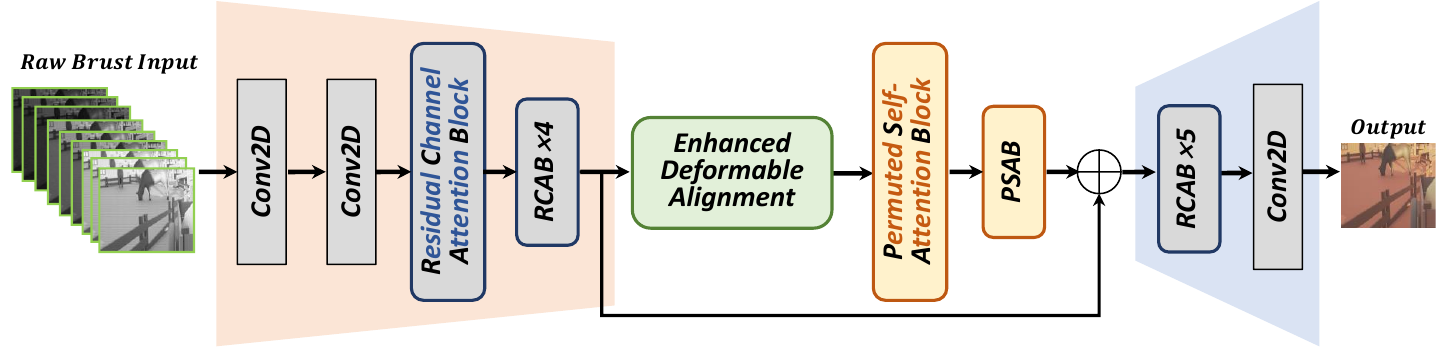}
    \caption{Our proposed enhanced alignment and permuted self-attention for burst HDR and restoration.}
    \label{fig:method}
\end{figure*}
 
\vspace{-4mm}
\paragraph{Training details.}  The proposed method was trained from scratch using only the challenge dataset. To train DarkIR-Fusion, the team used AdamW with setting $\beta_1=0.9$, $\beta_2=0.9$ and weight decay to $10^{-3}$. The learning rate was initialized in $10^{-3}$ and was updated following a cosine annealing strategy \cite{loshchilov2022sgdr} to a minimum $10^{-7}$. Rectangular crops of varying sizes (scale 2:1) were used, with random horizontal and vertical flips applied. The training follows a multi-step strategy, reducing the batch size in each step, while making the crop size larger. Across four different consecutive steps for [1500, 500, 300, 20] epochs, they used patch sizes [(256,512), (384,768), (512,1024), (768,1536)] and batch sizes [32, 16, 8, 4], respectively. The model was trained on four H100 GPUs approximately 20 hours.
The following combination of losses was used for optimization:
\vspace{-2mm}
\begin{equation} \label{eq: cidautai loss}
    \mathcal{L} = \lVert x - \hat{x} \rVert^2 + 50 \cdot \lVert \nabla x - \nabla \hat{x} \rVert^2 + 0.8 \cdot \lVert x_{\downarrow8} - \hat{x}_{\downarrow8} \rVert_{YUV}. 
\end{equation}

All distances in \eqref{eq: cidautai loss} were calculated using the $\mathcal{L}_1$ norm. $\nabla$ stands for gradient operation and YUV for the color space transformation. The reference ground-truth image and the cleaned image are, respectively, $x$ and $\hat{x}$.

\subsection{X-L}

\textbf{General method description.} In the task of burst HDR and restoration, efficiently extracting information from multiple frames and performing high-performance multi-frame alignment, fusion, and reconstruction is crucial. To address this, team X-L proposed a method using a feature encoder that includes five Residual Channel Attention Blocks (RCABs) to extract key information~\cite{zhang2018image}. These features are then fed into an enhanced deformable alignment module~\cite{luo2021ebsr}, which aligns neighboring frames to the reference frame. After the alignmentmodule, the features are processed through a cascaded Permuted Self-Attention Blocks (PSABs) to model global relationships~\cite{zhou2023srformer}. Finally, the features are fed to a decoder, which consists of five RCABs and one convolutional layer, to produce the final output. This approach ensures robust alignment and effective global context modeling.

\textbf{Training strategy.} During the training phase, they utilized the Adam optimizer with exponential decay rates set to $\beta_1 = 0.9$ and $\beta_2 = 0.999$. The initial learning rate was configured at $4 \times 10^{-4}$ and was halved every 200 epochs. 
They utilized the Charbonnier penalty loss, and the batch size was set to 4. 
The proposed method was implemented using the PyTorch framework and executed on a single NVIDIA 4090 GPU.

\section{Discussion and Conclusion} \label{sec: conclusion}
\subsection{Discussion}

Observing the diverse approaches of participants, it is first worth noting that almost all participants shared a common model structure, performing frame alignment first and then restoration. Indeed, as team ImvisionAI analyzed this competition in Section \ref{sec: imvisionai}, the approach of dividing the roles of each module and solving the problem step by step, rather than performing multiple tasks simultaneously with a single model, achieved effective results. 
Although there were small variations in how the optical flow was computed (e.g., Spynet\cite{ranjan2017optical} (team ImvisionAI), TMRnet\cite{jin2021temporal} (team DeepTrans), U-net\cite{ronneberger2015unet} (team SimonLarsen), simple deformable encoder\cite{dai2017deformable} (team X-L),, etc.), computing the optical flow to induce and influence warping took precedence over the restoration stage. Further note that, team SimonLarsen and team X-L tried multi-frame alignment on the extracted feature spaces. This approach would be significant to conduct additional studies to determine whether aligning over the denoised or extracted time-spatial features, as team X-L tried, could yield better results.

Secondly, most teams imposed a simple loss function, which is a common method in competitions to maximize the PSNR value. Team ImvisionAI used PSNR loss directly, while team DeepTrans and team SimonLarsen used the $L^1$ and $L^2$ loss functions, respectively. Team CidautAI further imposed an additional gradient-based loss \eqref{eq: cidautai loss}, but it still shares a similar insight with the $L^1$ loss. If one needs to solve tasks where qualitative results are more important and increasing PSNR is not the top priority, using additional losses and regularizations such as score distillation \cite{vincent2011connection, song2019generative, kim2022noise, nguyen2024swiftbrush} may lead to improved results.

Thirdly, we discuss the efficiency of the submitted models. Although team ImvisionAI's model records the best in performance, it also tightly consumes the competition's resource constraints. The second-place team, DeepTrans, achieved even better results than SimonLarsen with fewer parameters. Although it consumed more FLOPs, it was measured to have a faster inference time when provided with a high-end GPU. 
If the environment requires a low-end GPU or NPU and inference time is more important, such as in a mobile smartphone, team SimonLarsen's results may be more suitable.
Lastly, team CidautAI focused on optimizing the model and achieved the minimum inference time. Their approach could be developed further by slightly increasing the model size (simply the number of layers and channels) to achieve better performance, sacrificing a little inference time.


\subsection{Conclusion}
This paper provides a comprehensive overview of the \emph{NTIRE 2025 Efficient Burst HDR and Restoration} challenge and its results. Out of 217 registered participants, six teams submitted their final results. The competition introduced a novel Burst HDR fusion dataset consisting of 300 scenes, each with nine multi-exposure input frames. Team ImvisionAI emerged as the winner, achieving the highest PSNR value of 43.22 dB on this dataset. They introduced a recursive \emph{Flow-based Enhanced Deformable Alignment} (F-EDA) module for handling multi-exposure images and adopted a two-stage training strategy to separately learn frame alignment and restoration tasks.
The other five teams also proposed efficient models and learning strategies, each emphasizing different aspects of time and memory efficiency, which could benefit related tasks in the future.

Finally, we suggest that this competition could be further enhanced by considering additional metrics, such as both PSNR and runtime, or by introducing further degradations in image signal processing (e.g., bad pixel correction, fixed-pattern noise, lens shading correction, etc.).

\section*{Acknowledgments}
This work was partially supported by the Humboldt Foundation. We thank the NTIRE 2025 sponsors: ByteDance, Meituan, Kuaishou, and University of Wurzburg (Computer Vision Lab).

\appendix
\section{Teams and affiliations} \label{app: affiliation}

\newcommand{\team}[4]{
\subsection*{#1}
\textit{\textbf{Title:}} \\ {#2} \\
\textit{\textbf{Members:}} \\ {#3} \\
\textit{\textbf{Affiliations:}} \\{#4} \\
}

\team
{NTIRE 2025 team}
{NTIRE 2025 Efficient Burst HDR and Restoration Challenge}
{Sangmin Lee\textsuperscript{1} (\href{mailto:sanggmin.lee@samsung.com}{sanggmin.lee@samsung.com}),
Eunpil Park\textsuperscript{1}, 
Hyunhee Park\textsuperscript{1},
Hyung-Ju Chun\textsuperscript{1},
Youngjo Kim\textsuperscript{1},
Angel Canelo\textsuperscript{1}, 
Chongyi Li\textsuperscript{2}, 
ChunLe Guo\textsuperscript{2}, 
Xin Jin\textsuperscript{2}, 
Radu Timofte\textsuperscript{3}}
{
\textsuperscript{1} Department of Camera Innovation Group, Samsung Electronics, South Korea \\
\textsuperscript{2} College of Computer Science, Nankai University, China\\
\textsuperscript{3} Computer Vision Lab, University of W\"urzburg, Germany
}


\team
{ImvisionAI}
{Recursive Multi-Exposure Alignment with Spatiotemporal Decoupling for
Efficient Burst HDR and Restoration}
{
Qi Wu\textsuperscript{1} (\href{mailto:wuqiresearch@gmail.com}{wuqiresearch@gmail.com}), Tianheng Qiu\textsuperscript{2}, 
Yuchun Dong\textsuperscript{1},
Shenglin Ding\textsuperscript{1},
Guanghua Pan\textsuperscript{1}}
{
\textsuperscript{1} Imvision \\
\textsuperscript{2} University of Science and Technology of China, China}

\team
{DeepTrans}
{Flow-Guided Deformable Alignment with Channel-wise Self-Attention Reconstruct for Efficient Burst HDR Restoration}
{
Weiyu Zhou\textsuperscript{1} (\href{mailto:17355710941@163.com}{17355710941@163.com}), 
Tao Hu\textsuperscript{1},
Yixu Feng\textsuperscript{1},
Duwei Dai\textsuperscript{2},
Yu Cao\textsuperscript{3},
Peng Wu\textsuperscript{1},
Wei Dong\textsuperscript{4},
Yanning Zhang\textsuperscript{1},
Qingsen Yan\textsuperscript{1}
}
{
\textsuperscript{1} School of Computer Science, Northwestern Polytechnical University, China \\
\textsuperscript{2} Second Affiliated Hospital Of Xi’an Jiaotong University, China \\
\textsuperscript{3} Xi’an Institute of Optics and Precision Mechanics of CAS, China \\
\textsuperscript{4} Xi’an University of Architecture and Technology, China
}

\team
{SimonLarsen}
{Efficient Alignment and Fusion with Dense Flow Estimation and ConvNeXt Residual Dense Blocks}
{Simon J. Larsen\textsuperscript{1} (\href{mailto:sjl@esoft.com}{sjl@esoft.com}), 
}
{
\textsuperscript{1} Esoft, Denmark
}

\team
{E\_Group}
{A Hierarchical Fourier-based Transformer for Burst HDR and Restoration}
{
Ruixuan Jiang\textsuperscript{1} (\href{mailto:rxjiang21@gmail.com}{rxjiang21@gmail.com}), 
Senyan Xu\textsuperscript{1},
Xingbo Wang\textsuperscript{1},
Xin Lu\textsuperscript{1}
}
{
\textsuperscript{1} University of Science and Technology of China, China
}

\team
{CidautAI}
{Efficient Burst Aligning HDR Restoration based in FFT and Large Receptive Field Spatial Attention}
{
Marcos V. Conde\textsuperscript{1} (\href{mailto:marcos.conde@uni-wuerzburg.de}{marcos.conde@uni-wuerzburg.de}),
Javier Abad-Hernández\textsuperscript{1}, 
Álvaro García-Lara\textsuperscript{1}, 
Daniel Feijoo\textsuperscript{1}, 
Álvaro García\textsuperscript{1}
}
{
\textsuperscript{1} CidautAI
}

\team
{X-L}
{Burst HDR and Restoration via Enhanced Alignment and Permuted Self-Attention}
{Zeyu Xiao\textsuperscript{1} (\href{mailto:zeyuxiao1997@163.com}{zeyuxiao1997@163.com}), 
Zhuoyuan Li\textsuperscript{2}
}
{
\textsuperscript{1} National University of Singapore, Singapore \\
\textsuperscript{2} University of Science and Technology of China, China
}

{
\small
\bibliographystyle{ieeenat_fullname}

}

\end{document}